\documentclass[runningheads]{llncs}

\usepackage[utf8]{inputenc}

\usepackage{hyperref}

\usepackage{csquotes}

\usepackage{graphicx}
\graphicspath{{images/}}

\usepackage{xcolor}

\usepackage{array}

\usepackage{float}

\begin{document}
\title{Collecting Service-Based Maintainability Metrics from RESTful API Descriptions: Static Analysis and Threshold Derivation}

\titlerunning{Collecting Service-Based Maintainability Metrics from REST APIs}

\author{
	Justus Bogner\inst{1}\orcidID{0000-0001-5788-0991} \and
	Stefan Wagner\inst{1}\orcidID{0000-0002-5256-8429} \and
	Alfred Zimmermann\inst{2}\orcidID{0000-0003-3352-7207}
}

\authorrunning{J. Bogner et al.}

\institute{
    Institute of Software Engineering, University of Stuttgart, Germany\\
	\email{\{justus.bogner,stefan.wagner\}@iste.uni-stuttgart.de}	
	\and
	Herman Hollerith Center, University of Applied Sciences Reutlingen, Germany\\
	\email{alfred.zimmermann@reutlingen-university.de}
}

\maketitle

\begin{abstract}
While many maintainability metrics have been explicitly designed for service-based systems, tool-supported approaches to automatically collect these metrics are lacking.
Especially in the context of microservices, decentralization and technological heterogeneity may pose challenges for static analysis.
We therefore propose the modular and extensible RAMA approach (RESTful API Metric Analyzer) to calculate such metrics from machine-readable interface descriptions of RESTful services.
We also provide prototypical tool support, the RAMA CLI, which currently parses the formats OpenAPI, RAML, and WADL and calculates 10 structural service-based metrics proposed in scientific literature.
To make RAMA measurement results more actionable, we additionally designed a repeatable benchmark for quartile-based threshold ranges (green, yellow, orange, red).
In an exemplary run, we derived thresholds for all RAMA CLI metrics from the interface descriptions of 1,737 publicly available RESTful APIs.
Researchers and practitioners can use RAMA to evaluate the maintainability of RESTful services or to support the empirical evaluation of new service interface metrics.

\end{abstract}

\section{Introduction}
\label{sec:intro}
Maintainability, i.e. the degree of effectiveness and efficiency with which a software system can be modified to correct, improve, extend, or adapt it~\cite{ISO25010}, is an essential quality attribute for long-living software systems.
To manage and control maintainability, quantitative evaluation with metrics~\cite{Coleman1994} has long established itself as a frequently employed practice.
In systems based on service orientation~\cite{Papazoglou2003}, however, many source code metrics lose their importance due to the increased level of abstraction~\cite{Bogner2019-ICSME}.
For microservices as a lightweight and fine-grained service-oriented variant~\cite{Newman2015}, factors like the large number of small services, their decentralized nature, or high degree of technological heterogeneity may pose difficulties for metric collection and the applicability of existing metrics, which has also been reported in the area of performance testing~\cite{Eismann2020}.
Several researchers have therefore focused on adapting existing metrics and defining new metrics for service orientation (see e.g. our literature review~\cite{Bogner2017-IWSM} or the one from Daud and Kadir~\cite{Daud2014}).

However, approaches to automatically collect these metrics are lacking and for the few existing ones, tool support is rarely publicly available (see Section~\ref{sec:relWork}).
This significantly hinders empirical metric evaluation as well as industry adoption of service-based metrics.
To circumvent the described challenges, we therefore propose a metric collection approach focused on machine-readable RESTful API descriptions.
RESTful web services are resource-oriented services that employ the full HTTP protocol with methods like \texttt{GET}, \texttt{POST}, \texttt{PUT}, or \texttt{DELETE} as well as HTTP status codes to expose their functionality on the web~\cite{Pautasso2014}.
For microservices, RESTful HTTP is used as one of the primary communication protocols~\cite{Newman2015}.
Since this protocol is popular in industry~\cite{Schermann2016,Bogner2019-ICSA} and API documentation formats like WADL\footnote{\url{https://www.w3.org/submission/wadl}}, OpenAPI\footnote{\url{https://www.openapis.org}}, or RAML\footnote{\url{https://raml.org}} are widely used, such an approach should be broadly applicable to real-world RESTful services.
Relying on machine-readable RESTful documentation avoids having to implement tool support for several programming languages.
Second, such documents are often created reasonably early in the development process if a design-first approach is used.
And lastly, if such documents do not exist for the system, they can often be generated automatically, which is supported for popular RESTful frameworks like e.g. Spring Boot\footnote{\url{https://springdoc.org}}.

While formats like OpenAPI have been used in many analysis and reengineering approaches for service- and microservice-based systems~\cite{Neumann2018,Petrillo2018,Mayer2018}, there is so far no broadly applicable and conveniently extensible approach to calculate structural service-based maintainability metrics from interface specifications of RESTful services.
To fill this gap, we propose a new modular approach for the static analysis of RESTful API descriptions called RAMA (RESTful API Metric Analyzer), which we describe in Section~\ref{sec:approach}.
Our prototypical tool support to show the feasibility of this approach, the RAMA CLI, is able to parse the popular formats OpenAPI, RAML, and WADL and calculates a variety of service interface metrics related to maintainability.
Lastly, we also conducted a benchmark-based threshold derivation study for all metrics implemented in the RAMA CLI to make measurements more actionable for practitioners (see Section~\ref{sec:benchmark}).

\section{Related Work}
\label{sec:relWork}
Because static analysis for service orientation is very challenging, most proposals so far focused on programming language independent techniques.
In the context of service-oriented architecture (SOA), Gebhart and Abeck~\cite{Gebhart2011} developed an approach that extracts metrics from the UML profile SoaML (Service-oriented architecture Modeling Language).
The used metrics are related to the quality attributes unique categorization, loose coupling, discoverability, and autonomy.

For web services, several authors also used WSDL documents as the basis for maintainability evaluations.
Basci and Misra~\cite{Basci2009} calculated complexity metrics from them, while Sneed~\cite{Sneed2010} designed a tool-supported WSDL approach with metrics for quantity or complexity as well as maintainability design rules.

To identify linguistic antipatterns in RESTful interfaces, Palma et al.~\cite{Palma2015a} developed an approach that relies on semantic text analysis and algorithmic rule cards.
They do not use API descriptions like OpenAPI.
Instead, their tool support invokes all methods of an API under study to document the necessary information for the rule cards.

Finally, Haupt et al.~\cite{Haupt2017} published the most promising approach.
They used an internal canonical data model to represent the REST API and converted both OpenAPI and RAML into this format via the epsilon transformation language (ETL).
While this internal model is beneficial for extensibility, the chosen transformation relies on a complex model-driven approach.
Moreover, the extensibility for metrics remains unclear and some of the implemented metrics simply count structural attributes like the number of resources or the number of POST requests.
The model also does not take data types into account, which are part of many proposed service-based cohesion or complexity metrics.
So, while the general approach from Haupt et al. is a sound foundation, we adjusted it in several areas and made our new implementation publicly available.

\section{The RAMA Approach}
\label{sec:approach}
In this section, we present the details of our static analysis approach called RAMA (RESTful API Metric Analyzer).
To design RAMA, we first analyzed existing service-based metrics to understand which of them could be derived solely from service interface definitions and what data attributes would be necessary for this.
This analysis relied mostly on the results of our previous literature review~\cite{Bogner2017-IWSM}, but also took some newer or not covered publications into account.
Additionally, we analyzed existing approaches for WSDL and OpenAPI (see Section~\ref{sec:relWork}).
Based on this analysis, we then developed a data model, an architecture, and finally prototypical tool support.

\begin{figure}
    \centering
    \includegraphics[width=\linewidth]{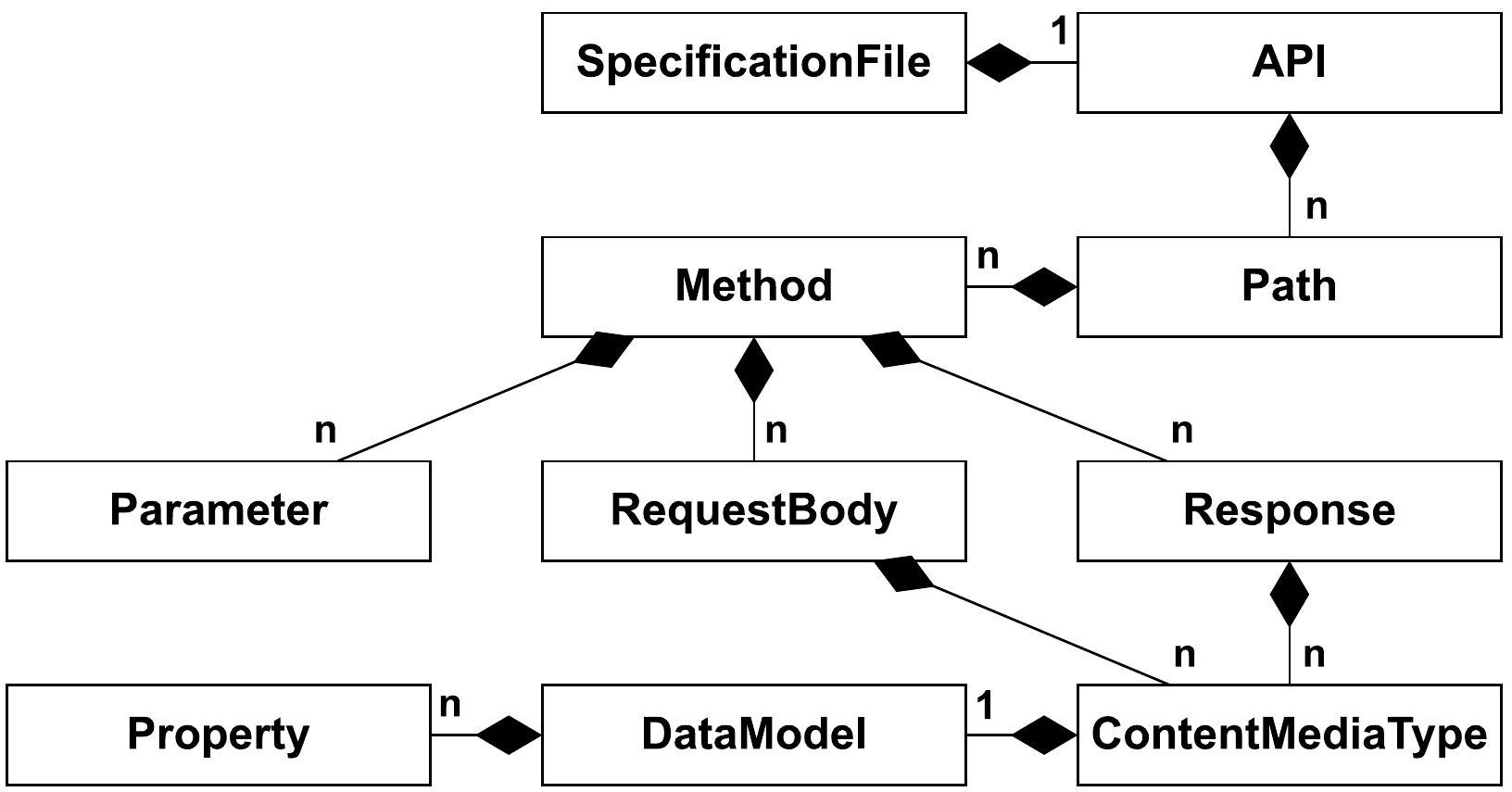}
    \caption{Simplified canonical data model of RAMA.}
    \label{fig:data-model-rest}
\end{figure}

Relying on a canonical data model to which each specification format has to be converted increases the independence and extensibility of our approach.
RAMA's internal data model (see Figure~\ref{fig:data-model-rest}) was constructed based on entities required to calculate a wide variety of complexity, size, and cohesion metrics.
While we tried to avoid unnecessary properties, we still needed to include all metric-relevant attributes and also to find common ground between the most popular RESTful description languages.

The hierarchical model starts with a \texttt{SpecificationFile} entity that contains necessary metadata like a title, a version, or the specification format (e.g. OpenAPI or RAML).
It also holds a single \texttt{API} wrapper entity consisting of a base path like e.g. \texttt{/api/v1} and a list of \texttt{Paths}.
These \texttt{Paths} are the actual REST resources of the API and each one of them holds a list of \texttt{Methods}.
A \texttt{Method} represents an HTTP verb like \texttt{GET} or \texttt{POST}, i.e. in combination, a \texttt{Path} and a \texttt{Method} form a service operation, e.g. \texttt{GET /customers/1/orders} to fetch all orders from customer with ID 1.
Additionally, a \texttt{Method} may have inputs, namely \texttt{Parameters} (e.g. path or query parameters) and \texttt{RequestBodies}, and outputs, namely \texttt{Responses}.
Since \texttt{RequestBodies} and \texttt{Responses} are usually complex objects of \texttt{ContentMediaTypes} like JSON or XML, they are both represented by a potentially nested \texttt{DataModel} with \texttt{Properties}.
Both \texttt{Parameters} and \texttt{Properties} contain the used data types, as this is important for cohesion and complexity metrics.
This model represents the core of the RAMA approach.

Based on the described data model, we designed the general architecture of RAMA as a simple command line interface (CLI) application that loosely follows the \textit{pipes and filters} architectural style.
One module type in this architecture is \texttt{Parser}.
A \texttt{Parser} takes a specific REST description language like OpenAPI as input and produces our canonical data model from it.
\texttt{Metrics} represent the second module type and are calculated from the produced data model.
The entirety of calculated \texttt{Metrics} form a summarized results model, which is subsequently presented as the final output by different \texttt{Exporters}.
This architecture is easily extensible and can also be embedded in other systems or a CI/CD pipeline.

\begin{figure}
    \centering
    \includegraphics[width=\linewidth]{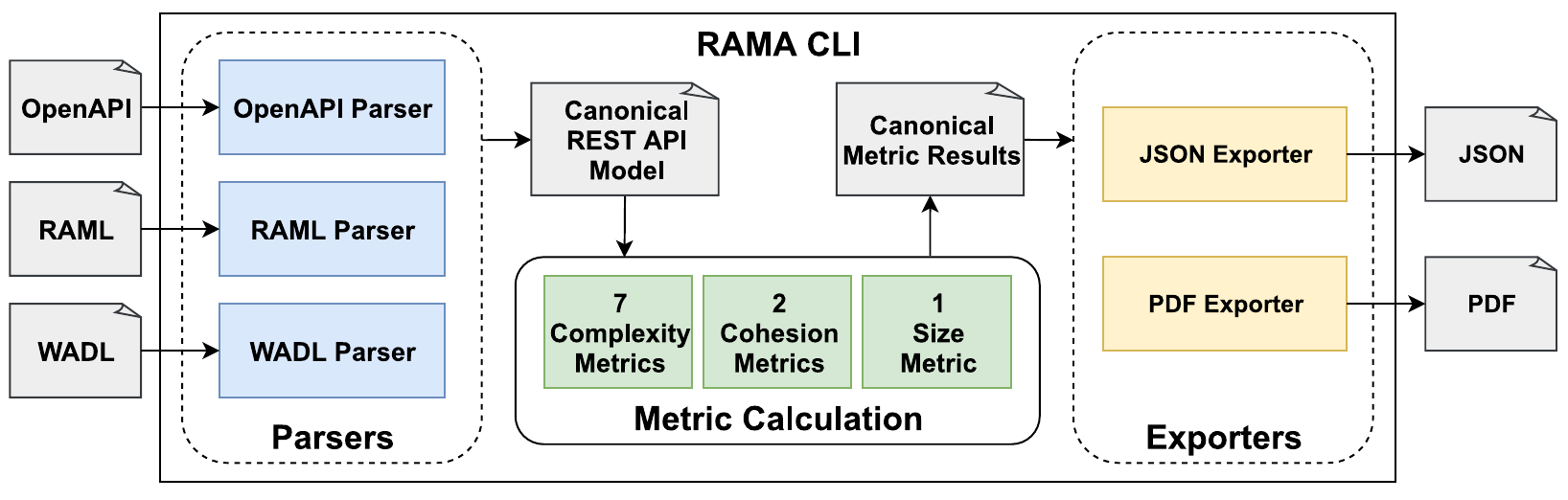}
    \caption{Implemented architecture of the RAMA CLI (arrows indicate data flow).}
    \label{fig:implementation-rest}
\end{figure}

The prototypical implementation of this approach is the RAMA CLI\footnote{\url{https://github.com/restful-ma/rama-cli}}.
It is written in Java and uses Maven for dependency management.
For metric modules, a plugin mechanism based on Java interfaces and the Java Reflection API enables the dynamic inclusion of newly developed metrics.
We present an overview of the implemented modules in Figure~\ref{fig:implementation-rest}.

For our internal data model, we used the \textit{protocol buffers} format\footnote{\url{https://developers.google.com/protocol-buffers}} developed by Google.
Since it is language- and platform-neutral and is easily serializable, it can be used in diverse languages and technologies.
There is also a tooling ecosystem around it that allows conversion between protocol buffers and various RESTful API description formats.
From this created \texttt{protobuf} model, the necessary Java classes are automatically generated (\texttt{Canonical REST API Model} in Figure~\ref{fig:implementation-rest}).

With respect to input formats, we implemented \texttt{Parsers} for OpenAPI, RAML, and WADL, since these are among the most popular ones based on GitHub stars, Google search hits, and StackOverflow posts~\cite{Haupt2018}.
Moreover, most of them offer a convenient tool ecosystem that we can use in our \texttt{Parser} implementations.
A promising fourth candidate was the Markdown-based API Blueprint\footnote{\url{https://apiblueprint.org}}, which seems to be rising in popularity.
However, since there is so far no Java parser for this format, we did not include it in the first prototype.

The RAMA CLI currently implements 10 service-based maintainability \texttt{Metrics} proposed in five different scientific publications (see Table~\ref{tab:rest-metrics}), namely seven complexity metrics, two cohesion metrics, and one size metric.
We chose these metrics to cover a diverse set of structural REST API attributes, which should demonstrate the potential scope of the approach.
We slightly adjusted some of the metrics for REST, e.g. the ones proposed for WSDL.
For additional details on each metric, please refer to our documentation\footnote{\url{https://github.com/restful-ma/rama-cli/tree/master/docs/metrics}} or the respective source.

Finally, we implemented two \texttt{Exporters} for the CLI, namely one for a PDF and one for a JSON file.
Additionally, the CLI automatically outputs the results to the terminal.
While this prototype already offers a fair amount of features and should be broadly applicable, the goal was also to ensure that it can be extended with little effort.
In this sense, the module system and the usage of interfaces and the Reflection API make it easy to add new \texttt{Parsers}, \texttt{Metrics}, or \texttt{Exporters} so that the RAMA CLI can be of even more value to practitioners and researchers.

\begin{table}
	\centering
	\caption{Implemented maintainability metrics of the RAMA CLI.}
	\label{tab:rest-metrics}
	\begin{tabular}{lllll}
		Name & Abbrev. & Property & Source\\
		\hline
        \hline
        Average Path Length & APL & Complexity & Haupt et al.~\cite{Haupt2017}\\
        Arguments per Operation & APO & Complexity & Basci and Misra~\cite{Basci2009}\\
        Biggest Root Coverage & BRC & Complexity & Haupt et al.~\cite{Haupt2017}\\
        Data Weight & DW & Complexity & Basci and Misra~\cite{Basci2009}\\
        Distinct Message Ratio & DMR & Complexity & Basci and Misra~\cite{Basci2009}\\
        Longest Path & LP & Complexity & Haupt et al.~\cite{Haupt2017}\\
        Number of Roots & NOR & Complexity & Haupt et al.~\cite{Haupt2017}\\
        Lack of Message-Level Cohesion & LoC$_{msg}$ & Cohesion & Athanasopoulos et al.~\cite{Athanasopoulos2015}\\
        Service Interface Data Cohesion & SIDC & Cohesion & Perepletchikov et al.~\cite{Perepletchikov2007}\\
		Weighted Service Interface Count & WSIC & Size & Hirzalla et al.~\cite{Hirzalla2009}\\
	\end{tabular}
\end{table}

\section{Threshold Benchmarking}
\label{sec:benchmark}
Metric values on their own are often difficult to interpret.
Some metrics may have a lower or an upper bound (e.g. a percentage between 0 and 1) and may also indicate that e.g. lower values are better or worse.
However, that is often still not enough to derive implications from a specific measurement.
To make metric values more actionable, \textit{thresholds} can therefore play a valuable role~\cite{Vale2019}.
We therefore designed a simple, repeatable, and adjustable threshold derivation approach to ease the application of the metrics implemented within RAMA.

\subsection{Research Design}
Since it is very difficult to rigorously evaluate a single threshold value, the majority of proposed threshold derivation methods analyze the measurement distribution over a large number of real-world systems.
These methods are called benchmark-based approaches~\cite{Baggen2012} or portfolio-based approaches~\cite{Brauer2017}.
Since a large number of RESTful API descriptions are publicly available, we decided to implement a simple benchmark-based approach.

Inspired by Bräuer et al.~\cite{Brauer2017}, we formed our labels based on the quartile distribution.
Therefore, we defined a total of four ranked bands into which a metric value could fall (see also Table~\ref{tab:rest-thresholds}), i.e. with the derived thresholds, a measurement could be in the top 25\%, between 25\% and the median, between the median and 75\%, or in the bottom 25\%.
Depending on whether lower is better or worse for the metric, each band was associated with one of the colors green, yellow, orange, and red (ordered from best to worst).
If a metric result is in the worst 25\% (red) or between the median and the worst 25\% (orange) of analyzed systems, it may be advisable to improve the related design property.

\begin{table}
	\centering
	\caption{Used metric threshold bands (colors are based on a metric where lower is better; for metrics where higher is better, the color ordering would be reversed).}
	\label{tab:rest-thresholds}
	\begin{tabular}{llll}
		Band & Color & Start & End\\
		\hline
        \hline
        Q1 & \textcolor{green}{\rule{0.3cm}{.3cm}} green & lower bound or minimum & 1$^{st}$ quartile\\
        Q2 & \textcolor{yellow}{\rule{0.3cm}{.3cm}} yellow & 1$^{st}$ quartile & 2$^{nd}$ quartile / median\\
        Q3 & \textcolor{orange}{\rule{0.3cm}{.3cm}} orange & 2$^{nd}$ quartile / median & 3$^{rd}$ quartile\\
        Q4 & \textcolor{red}{\rule{0.3cm}{.3cm}} red & 3$^{rd}$ quartile & upper bound or maximum\\
	\end{tabular}
\end{table}

\begin{figure}
    \centering
    \includegraphics[width=\linewidth]{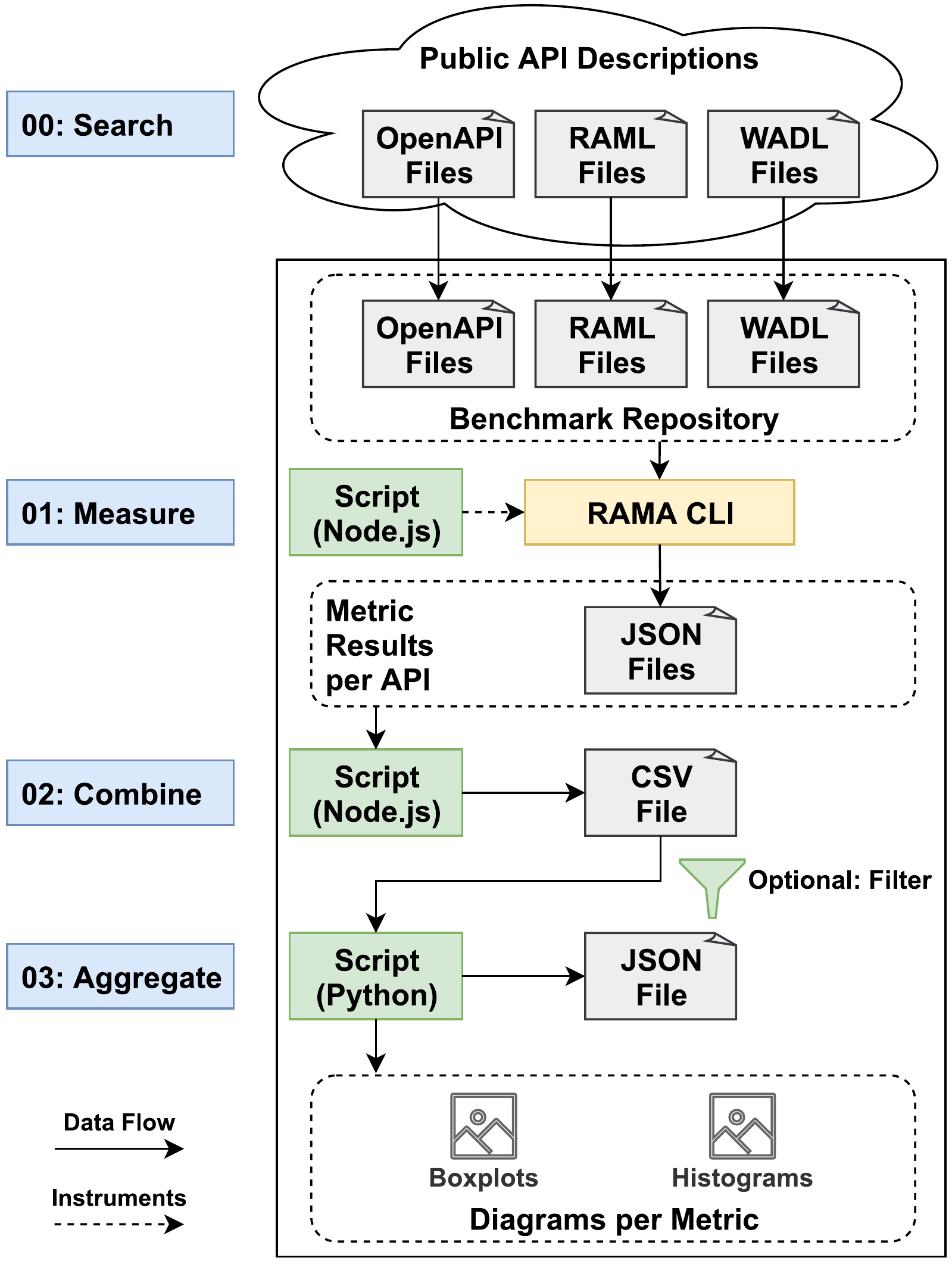}
    \caption{Threshold benchmark design.}
    \label{fig:benchmark}
\end{figure}

To derive these thresholds per RAMA CLI metric, we designed an automated benchmark pipeline that operates on a large number of API description files.
The benchmark consists of the four steps \texttt{Search}, \texttt{Measure}, \texttt{Combine}, and \texttt{Aggregate} (see Figure~\ref{fig:benchmark}).
The first step was to search for publicly available descriptions of real-world APIs.
For this, we used the keyword and file type search on GitHub.
Additionally, we searched the API repository from APIs.guru\footnote{\url{https://apis.guru/browse-apis}}, which provides a substantial number of OpenAPI files.

Once a sufficiently large collection of parsable files had been established, we collected the metrics from them via the RAMA CLI (\texttt{Measure} step).
In the third step \texttt{Combine}, this collection of JSON files was then analyzed by a script that combined them into a single CSV file, where each analyzed API represented a row.
Using this file with all measurements, another script executed the threshold analysis and aggregation (\texttt{Aggregate} step). Optionally, this script could filter out APIs, e.g. too small ones.
As results, this yielded a JSON file with all descriptive statistics necessary for the metric thresholds as well as two diagram types to potentially analyze the metric distribution further, namely a histogram and a boxplot, both in PNG format.

To make the benchmark as transparent and repeatable as possible, we published all related artifacts such as scripts, the used API files, and documentation in a GitHub repository\footnote{\url{https://github.com/restful-ma/thresholds}}.
Every subsequent step after \texttt{Search} is fully automatable and we also provide a wrapper script to execute the complete benchmark with one command.
Our goal is to provide a reusable and adaptable foundation for re-executing this benchmark with different APIs as input that may be more relevant threshold indicators for a specific REST API under analysis.

\subsection{Results}
We initially collected 2,651 real-world API description files (2,619 OpenAPI, 18 WADL, and 14 RAML files).
This sample was dominated by large cloud providers like Microsoft Azure (1,548 files), Google (305 files), or Amazon Web Services (205 files).
Additionally, there were cases where we had several files of different versions for the same API.

A preliminary analysis of the collected APIs revealed that a large portion of them were very small, with only two or three operations.
Since it seems reasonable to assume that several of the RAMA CLI metrics are correlated with size, we decided to exclude APIs with less than five operations (\textit{Weighted Service Interface Count} $<$ 5) to avoid skewing the thresholds in favor of very small APIs.
Therefore, we did not include 914 APIs in the \texttt{Aggregate} step.
Our exemplary execution of the described benchmark calculated the quartile-based thresholds based on a total of \textbf{1,737 public APIs} (1,708 OpenAPI, 16 WADL, and 13 RAML files).
The median number of operations for these APIs was 15.
Table~\ref{tab:metric-thresholds} lists the thresholds for all 10 metrics of the RAMA CLI.
Because of the sequential parsing of API files, the execution of the benchmark can take up to several hours on machines with low computing power.
We therefore also provide all result artifacts of this exemplary run in our repository\footnote{\url{https://github.com/restful-ma/thresholds/tree/master/results}}.

\begin{table}
    \centering
	\caption{Calculated metric thresholds from 1,737 API description files.}
	\label{tab:metric-thresholds}
	\begin{tabular}{lllll}
		Metric & \textcolor{green}{\rule{0.3cm}{.3cm}} Top 25\% & \textcolor{yellow}{\rule{0.3cm}{.3cm}} 25\% - 50\% & \textcolor{orange}{\rule{0.3cm}{.3cm}} 50\% - 75\% & \textcolor{red}{\rule{0.3cm}{.3cm}} Worst 25\%\\
		\hline
        \hline
        APO & [0.20, 3.52]  & ]3.52, 4.60] & ]4.60, 8.63] & ]8.63, 21.63]\\
        APL & [1.00, 2.50]  & ]2.50, 5.00] & ]5.00, 8.00] & ]8.00, 15.60]\\
        BRC & [1.00, 1.00]  & ]1.00, 0.99] & ]0.99, 0.60] & ]0.60, 0.00]\\
        DW & [4, 77]  & ]77, 167] & ]167, 378] & ]378, 41570]\\
        DMR & [0.00, 0.26]  & ]0.26, 0.36] & ]0.36, 0.48] & ]0.48, 1.00]\\
        LoC$_{msg}$ & [0.00, 0.53]  & ]0.53, 0.62] & ]0.62, 0.69] & ]0.69, 1.00]\\
        LP & [1, 3]  & ]3, 8] & ]8, 10] & ]10, 19]\\
        NOR & [1, 1]  & ]1, 2] & ]2, 3] & ]3, 359]\\
        SIDC & [1.00, 1.00]  & ]1.00, 0.64] & ]0.64, 0.55] & ]0.55, 0.00]\\
        WSIC & [5, 8]  & ]8, 15] & ]15, 31] & ]31, 1126]\\
	\end{tabular}
\end{table}

\section{Limitations and Threats to Validity}
While we pointed out several advantages of the RAMA approach, there are also some limitations.
First, RAMA only supports RESTful HTTP and therefore excludes asynchronous message-based communication.
Even though REST is arguably still more popular for microservice-based systems, event-driven microservices based on messaging receive more and more attention.
Similar documentation standards for messaging are slowly emerging (see e.g. AsyncAPI\footnote{\url{https://www.asyncapi.com}}), but our current internal model and metric implementations are very REST-specific.
While several metrics are undoubtedly valid in both communication paradigms, substantial efforts would be necessary to fully support messaging in addition to REST.
Apart from that, the approach requires machine-readable RESTful API descriptions to work.
While such specifications are popular in the RESTful world, not every service under analysis will have one.
And thirdly, relying on an API description file restricts the scope of the evaluation.
Collected metrics are focused on the interface quality of a single service and cannot make any statement about the concrete service implementation.
Therefore, RAMA cannot calculate system-wide metrics except for aggregates like mean, which also excludes metrics for the coupling between services.

Our prototypical implementation, the RAMA CLI, may also suffer from potential limitations.
While we tried to make it applicable to a wide range of RESTful services by supporting the three formats OpenAPI, RAML, and WADL, there are still other used formats for which we currently do not have a parser, e.g. API Blueprint\footnote{\url{https://apiblueprint.org}}.
Similarly, there are many more proposed service-based metrics we could have implemented in the RAMA CLI.
The modular architecture of RAMA consciously supports possible future extensions in this regard.
Lastly, we unfortunately cannot guarantee that the prototype is completely free of bugs and works reliably with every single specification file.
While we were very diligent during the implementation, have a test coverage of $\sim$75\%, and successfully used the RAMA CLI with over 2,500 API specification files, it remains a research prototype.
For transparency, the code is publicly available as open source and we welcome contributions like issues or pull requests.

Finally, we need to mention threats to validity concerning our empirical threshold derivation study.
One issue is that the derived thresholds rely entirely on the quality and relevance of the used API description files.
If the majority of files in the benchmark are of low quality, the derived thresholds will not be strict enough.
Measurement values of an API may then all fall into the Q1 band, when, in reality, the service interface under analysis is still not well designed.
By including a large number of APIs from trustworthy sources, this risk may be reduced.
However, there still may be services from specific contexts that are so different that they need a custom benchmark to produce relevant thresholds.
Examples could be benchmarks based only on a particular domain (e.g. cloud management), on a single API specification format (e.g. RAML), or on APIs of a specific size (e.g. small APIs with 10 or less operations).
As an example, large cloud providers like Azure, Google, or AWS heavily influenced our benchmark run.
Each one of these uses fairly homogeneous API design, which influenced some metric distributions and thresholds.
We also eliminated a large number of very small services with less than five operations to not skew metrics in this direction.
So, while our provided thresholds may be useful for a quick initial quality comparison, it may be sensible to select the input APIs more strictly to create a more appropriate size- or domain-specific benchmark.
To enable such replication, our benchmark focuses on repeatability and adaptability.

\section{Conclusion}
To support static analysis based on proposed service-based maintainability metrics in the context of microservices, we designed a tool-supported approach called RAMA (RESTful API Metric Analyzer).
Service interface metrics are collected based on machine-readable descriptions of RESTful APIs.
Our implemented prototypical tool, the RAMA CLI, currently supports the specification formats OpenAPI, RAML, and WADL as well as 10 metrics (seven for complexity, two for cohesion, and one size metric).
To aid with results interpretation, we also conducted an empirical benchmark that calculated quartile-based threshold ranges (green, yellow, orange, red) for all RAMA CLI metrics using 1,737 public RESTful APIs.
Since the thresholds are very dependent on the quality and relevance of the used APIs, we designed the automated benchmark to be repeatable.
Accordingly, we published the RAMA CLI\footnote{\url{https://github.com/restful-ma/rama-cli}} as well as all results and artifacts of the threshold derivation study\footnote{\url{https://github.com/restful-ma/thresholds}} on GitHub.

RAMA can be used by researchers and practitioners to efficiently calculate suitable service interface metrics for size, cohesion, or complexity, both for early quality evaluation or within continuous quality assurance.
Concerning possible future work, a straight-forward option would be the extension of the RAMA CLI with additional input formats and metrics to increase its applicability and utility.
Additionally, our static approach could be combined with existing dynamic approaches~\cite{Engel2018,Bogner2019-PROFES} to mitigate some of its described limitations.
However, the most critical expansion for this line of research is the empirical evaluation of proposed service-based maintainability metrics, as most authors did not provide such evidence.
Due to the lack of automatic collection approaches, such evaluation studies were previously challenging to execute at scale.
Our preliminary work can therefore serve as a valuable foundation for such endeavors.

\subsubsection*{Acknowledgments}
We kindly thank Marvin Tiedtke, Kim Truong, and Matthias Winterstetter for their help with the threshold study execution and tool development.
Similarly, we thank Kai Chen and Florian Grotepass for their implementation support.
This research was partially funded by the Ministry of Science of Baden-Württemberg, Germany, for the doctoral program \textit{Services Computing}\footnote{\url{https://www.services-computing.de/?lang=en}}.

\bibliographystyle{splncs04}
\bibliography{sources}

\end{document}